\documentclass[structabstract]{aa}
\usepackage{amssymb}
\usepackage{graphicx}
\usepackage{txfonts}
\begin{document}
   \title{The optical variability of steep-spectrum radio quasars in the SDSS Stripe 82 region}


   \author{Minfeng Gu
          \inst{1}
          \and
          Y. L. Ai\inst{2,3}
          }

   \institute{Key Laboratory for Research in Galaxies and Cosmology, Shanghai Astronomical Observatory,
    Chinese Academy of Sciences, 80 Nandan Road, Shanghai 200030, China\\
              \email{gumf@shao.ac.cn}
         \and National Astronomical Observatories/Yunnan Observatory,
             Chinese Academy of Sciences, P.O. Box 110, 650011 Kunming, Yunnan, China
         \and
             Key Laboratory for the Structure and Evolution of Celestial Objects,
             Chinese Academy of Sciences, P.O. Box 110, 650011 Kunming, Yunnan, China
             }



\titlerunning{Optical variability of SSRQs in Stripe 82}
\authorrunning{M. F. Gu \& Y. L. Ai}

  \abstract
   {While there are a lot of investigations on the optical variability of flat-spectrum radio quasars (FSRQs), 
   not much work has been done on the optical variability of steep-spectrum radio quasars (SSRQs).}
   {We investigate the optical variability in SSRQs. For comparison, the optical variability of 
   FSRQs are also explored.}
   {We investigate the optical variability of 18 SSRQs and 15 FSRQs in the SDSS
Stripe 82 region using SDSS DR7 released multi-epoch data covering
about nine years. We determined the spectral index by fitting a
powerlaw to SDSS ${ugriz}$ photometric data, and explored the
relationship between the spectral index and source brightness.}
   {For all SSRQs studied, we detect variations in $r$ band flux of overall amplitude between 0.22
mag and 0.92 mag in different sources. Eight of 18 SSRQs display
a bluer-when-brighter (BWB) trend. In contrast, the variability amplitude of 15 FSRQs in r band ranges from 0.18 to 0.97 mag, 
and 11 of 15 sources show a BWB trend.
We found an anti-correlation between the Eddington ratio and the variability amplitude 
in r band for SSRQs, which is similar to that in radio quiet AGNs. This implies that the thermal 
emission from accretion disk may be responsible for the variability in SSRQs, although the jet nonthermal 
emission cannot be excluded. In contrast, there is no correlation in FSRQs, which implies that the mechanisms of 
variability in FSRQs may be different from that in SSRQs. The optical continuum variability of radio loud
broad absorption line quasars (BALQs) are investigated for the first time here on two sources with steep
radio spectrum. Both radio loud BALQs show variations with amplitude of about 0.3 mag
at r band. Their spectral variability all show a BWB trend. 
In a combined sample (18 SSRQs, and 44 FSRQs) of our sample with the FSRQs in Gu \& Ai (2011), we found a trend of a broader line width of
broad Mg II emission line with steeper radio spectral index. It implies a disc-like broad line region (BLR) geometry may present 
in these quasars. In these 62 quasars, we found a $\sim57\%$ source percentage showing BWB trend in FSRQs, whereas it 
is $\sim44\%$ in SSRQs.}
   {}

   \keywords{galaxies: active -- galaxies: quasars: general -- galaxies: photometry}

   \maketitle
%

\section{Introduction}

Active galactic nuclei (AGNs) are characteristic of variability at 
almost all wavelength (e.g. Wiita 1996). Multiwavelength studies of
variations of AGNs have played important roles in exploring the physical conditions
near the center of AGNs. Based on the 
variability, the reverberation mapping method was developed to calculate
the size of broad line region (BLR). From the photoionization model, 
the flux variations of BLR follows the flux variation of continuum ionization 
emission with certain time lag, which corresponds to the size of BLR
(Peterson 1993).  Moreover, the disc-jet connections can be investigated
by using the long term multiwavelength variability monitorings (e.g. Chatterjee et al. 2009, 2011).

Intrinsically, the variations of AGNs are generally caused by the physical variations in jet and accretion disc.
However, the contribution of each component in the observed variability varies from source to source. 
Blazars, including BL Lac objects and flat-spectrum radio quasars
(FSRQs), are the most extreme class of active galactic nuclei
(AGNs), characterized by strong and rapid variability, high
polarization, and apparent superluminal motion. These extreme
properties are generally interpreted as a consequence of non-thermal
emission from a relativistic jet oriented close to the line of
sight. In general, the variations in blazars are dominated by the jet emission.
There are extensive explorations on the optical variability of blazars 
(e.g. Ghisellini et al. 1997; Fan et al. 1998; Massaro et al.
1998; Ghosh et al. 2000; Clements \& Carini 2001; Raiteri et al.
2001; Villata et al. 2002; Vagnetti et al. 2003; Wu et al 2005,
2007; Gu et al. 2006; Hu et al. 2006; Poon et al. 2009; Rani et al. 2010; 
Gu \& Ai 2011). While it is generally accepted that the nonthermal emission 
from relativistic jet oriented close to the line of sight dominate the optical continuum, the 
situation seems more complicated in FSRQs. It is not clear whether the nonthermal 
jet emission plays the main role in the optical variability. Evidence of thermal emission 
in FSRQs has been detected in several cases usually during 
low activity states (e.g. 3C279, Pian et al. 1999; 3C273, Grandi \& Palumbo 2004; 
3C454.3, Raiteri et al. 2007). A recent work shows that the optical variability in the 
FSRQ PKS 1510-089 is mainly due to the thermal emission, and only during major flares a 
contribution from the jet is seen (D'Ammando et al. 2011).  Moreover, the redder-when-brighter trend 
found in several FSRQs implies that the thermal emission plays important role in the 
optical variability (e.g. Gu et al. 2006; Rani et al. 2010; Gu \& Ai 2011). 

Radio-quiet AGNs has also been explored in terms of the long-term or
short-term variability (e.g.  Stalin et al. 2004, 2005; Gupta \& Joshi 2005; 
Ai et al. 2010). The accretion disc instabilities may likely explain the 
optical microvariability in radio quiet AGNs (e.g. Gopal-Krishna et al. 1995, see also Gopal-Krishna et al. 2003 
for alternative scenario). In terms of the long-term variability, the change of accretion rate is used to explain the optical
variations in radio quiet AGNs (e.g. Li \& Cao 2008). The properties of 
steep spectrum radio quasars (SSRQs) are intermediate between FSRQs and radio
quiet quasars. SSRQs are usually lobe-dominated radio quasars, with radio 
lobe emission dominate over the radio core emission. Their jets are viewed at
larger angles than blazars. Therefore, the beaming effects from jets should not be 
severe (see e.g. Liu et al. 2006), and the jet emission is not expected 
to dominate at optical bands (e.g. Gu \& Ai 2011). However, the variations in SSRQs
could still be a mixture of being caused by jet and accretion disc, since they are powerful radio emitter.
The optical variations of SSRQs are largely unknown, and there are only few explorations on variations in SSRQs (e.g. Stalin et al. 2004, 2005).
While the color variations and/or spectral variations of blazars are extensively investigated 
(e.g. Ghisellini et al. 1997; Fan et al. 1998; Massaro et al.
1998; Ghosh et al. 2000; Clements \& Carini 2001; Raiteri et al.
2001; Villata et al. 2002; Vagnetti et al. 2003; Wu et al 2005,
2007; Gu et al. 2006; Rani et al. 2010; Gu \& Ai 2011), it has rarely been
done for SSRQs. In this work, we investigate the optical variability for a sample of
SSRQs, as well as their spectral variations.

The layout of this paper is as follows: in Section 2, we describe
the source sample; the variability results are outlined in Section
3; Section 4 includes the discussion; and in the last section, we
draw our conclusions. The cosmological parameters $H_{\rm 0}=70\rm~
km~ s^{-1}~ Mpc^{-1}$, $\Omega_{\rm m}=0.3$, and
$\Omega_{\Lambda}=0.7$ are used throughout the paper, and the
spectral index $\alpha$ is defined as $f_{\nu}\propto\nu^{-\alpha}$, where
$f_{\nu}$ is the flux density at frequency $\nu$.


\section{Sample selection}

\subsection{Quasars in Stripe 82 region}

Our initial quasar sample was selected as those quasars both in the
SDSS DR7 quasar catalogue (Schneider et al. 2010) and Stripe 82
region. The SDSS DR7 quasar catalogue consists of 105,783
spectroscopically confirmed quasars with luminosities brighter than
$M_{i}=-22.0$, with at least one emission line having a full width
at half-maximum (FWHM) larger than 1000 $\rm km~ s^{-1}$ and highly
reliable redshifts. The sky coverage of the sample is about 9380
$\rm deg^2$ and the redshifts range from 0.065 to 5.46. The
five-band $(u,~ g,~ r,~ i,~ z)$ magnitudes have typical errors of
about 0.03 mag. The spectra cover the wavelength range from 3800 to
9200 $\rm \AA$ with a resolution of $\simeq2000$ (see Schneider et
al. 2010 for details). The Stripe 82 region, i.e. right ascension
$\alpha = 20^{\rm h} - 4^{\rm h}$ and declination
$\delta=-1^{\circ}.25 - +1^{\circ}.25$, was repeatedly scanned
during the SDSS-I phase (2000 - 2005) under generally photometric
conditions, and the data are well calibrated (Lupton et al. 2002).
This region was also scanned repeatedly over the course of three
3-month campaigns in three successive years in 2005 - 2007 known as
the SDSS Supernova Survey (SN survey). The multi-epoch photometric
observations therefore enable us to investigate the optical
variability of the selected quasars.

\subsection{Cross-correlation with radio catalogues}

In this paper, we define a quasar to be a SSRQ according to its
radio spectral index. Therefore, we cross-correlate the initial
quasar sample with the Faint Images of the Radio Sky at Twenty
centimeters (FIRST) 1.4-GHz radio catalogue (Becker, White \&
Helfand 1995), the Green Bank 6-cm (GB6) survey at 4.85 GHz radio
catalogue (Gregory et al. 1996), and the Parkes-MIT-NRAO (PMN) radio
continuum survey at 4.85 GHz (Griffith \& Wright, 1993), as well as
for sources with $\delta<0^{\circ}$. The FIRST survey used the Very
Large Array (VLA) to observe the sky at 20 cm (1.4 GHz) with a beam
size of 5.4 arcsec. FIRST was designed to cover the same region of
the sky as the SDSS, and observed 9000 $\rm deg^2$ at the north
Galactic cap and a smaller $2^{\circ}.5$ wide strip along the
celestial equator. It is 95 per cent complete to 2 mJy and 80 per
cent complete to the survey limit of 1 mJy. The survey contains over
800,000 unique sources, with an astrometric uncertainty of $\lesssim
1$ arcsec.

The GB6 survey at 4.85 GHz was executed with the 91-m Green Bank
telescope in 1986 November and 1987 October. Data from both epochs
were assembled into a survey covering the
$0^{\circ}<\delta<75^{\circ}$ sky down to a limiting flux of 18 mJy,
with 3.5 arcmin resolution. GB6 contains over 75,000 sources, and
has a positional uncertainty of about 10 arcsec at the bright end
and about 50 arcsec for faint sources (Kimball \& Ivezi\'{c} 2008).
The PMN surveys were made using the Parkes 64-m radio telescope at a
frequency of 4850 MHz with the NRAO multibeam receiver mounted at
the prime focus (Griffith \& Wright 1993). The surveys had a spatial
resolution of approximately $4'.2$ FWHM and were made for the
southern sky between declinations of $-87^{\circ}$ and
$+10^{\circ}$, and all right ascensions during June and November in
1990. The positional accuracy is close to 10 arcsec in each
coordinate. The survey was divided into four declination bands. One
of these four is the equatorial survey
($-9^{\circ}.5<\delta<+10^{\circ}.0$) covering 1.90 sr, which
contains 11,774 sources to a flux limit of 40 mJy and largely
overlaps the GB6 survey in the declination range from $0^{\circ}$ to
$+10^{\circ}$ (Griffith et al. 1995).

The initial quasar sample was first cross-correlated between the
SDSS quasar positions and the FIRST catalogue to within 2 arcsec
(see e.g. Ivezi\'{c} et al. 2002; Lu et al. 2007). The resulting
sample of SDSS quasar positions was then cross-correlated with both
the GB6 and PMN equatorial catalogues to within 1 arcmin (e.g.
Kimball \& Ivezi\'{c} 2008). Owing to the different spatial
resolutions of FIRST, GB6, and PMN, multiple FIRST counterparts were
found to within 1 arcmin for some quasars, although there is only
single GB6 and/or PMN counterpart existed. The optical variability
of quasars with single FIRST counterparts to within 1 arcmin of the
SDSS positions has been presented in Gu \& Ai (2011). In this paper,
we focus on the optical variability of quasars with multiple FIRST
counterparts.

There are total 29 sources having multi FIRST counterparts within 1
arcmin of SDSS positions. The radio spectral index $\alpha_{\rm r}$ was then
calculated between the integrated FIRST 1.4 GHz from multi-counterparts
within 1 arcmin and either or both of the GB6 and PMN 4.85 GHz.
Eleven sources are defined as SSRQs with $\alpha_{\rm r}>0.5$ using the
integrated 1.4 GHz flux density. In comparison, only two sources are
defined as SSRQs if only using 1.4 GHz flux density of the closest
counterpart. This implies that most genuine SSRQs may appear as
multi-FIRST components, but be identified as FSRQs if using the flux
density of the closest FIRST component, which is actually much
smaller than that of radio lobes. When counterparts were found in
both the GB6 and PMN surveys, the spectral indices were consistent
with each other. To further check the spectral index, we searched for the
counterparts in  NRAO VLA Sky Survey (NVSS) within 1 arcmin of SDSS positions. 
The NVSS was also carried out using the VLA
at 1.4 GHz to survey the entire sky north of $\delta=-40^{\circ}$
and contains over 1.8 million unique detections brighter than 2.5
mJy, however with lower spatial resolution $\rm 45^{''} beam^{-1}$. 
Due to the lower spatial resolution, all sources have single NVSS
counterpart, except for five sources, which have two counterparts,
i.e. SDSS J015105.80$-$003426.4, SDSS J213513.10$-$005243.8, SDSS
J221409.96$+$005227.0, SDSS J220719.77$+$004157.3, and SDSS
J233624.04$+$000246.0. We then re-calculated the spectral index using
the flux density from single or integrated NVSS counterparts. We found that 
25 out of 29 sources have consistent spectral index
between using FIRST and NVSS. Three sources (SDSS J005905.51$+$000651.6, SDSS
J013352.66$+$011345.1 and SDSS J021225.56$+$010056.1) changed from
FSRQs to SSRQs using NVSS flux density, for which the part of
emission can be likely missed in FIRST images. One souce, i.e. SDSS
J015509.00$+$011522.5, changes from steep to flat spectral index,
but the spectral index are all close to 0.5. Therefore, it is still
treated as SSRQ. Among the 29 sources, 14 are defined as SSRQs with
$\alpha_{\rm r}\geq0.5$, whereas 15 objects are FSRQs with
$\alpha_{\rm r}<0.5$ (see Table \ref{table_source}).


\subsection{Sample}

\begin{figure}
   \centering
   \includegraphics[width=0.5\textwidth]{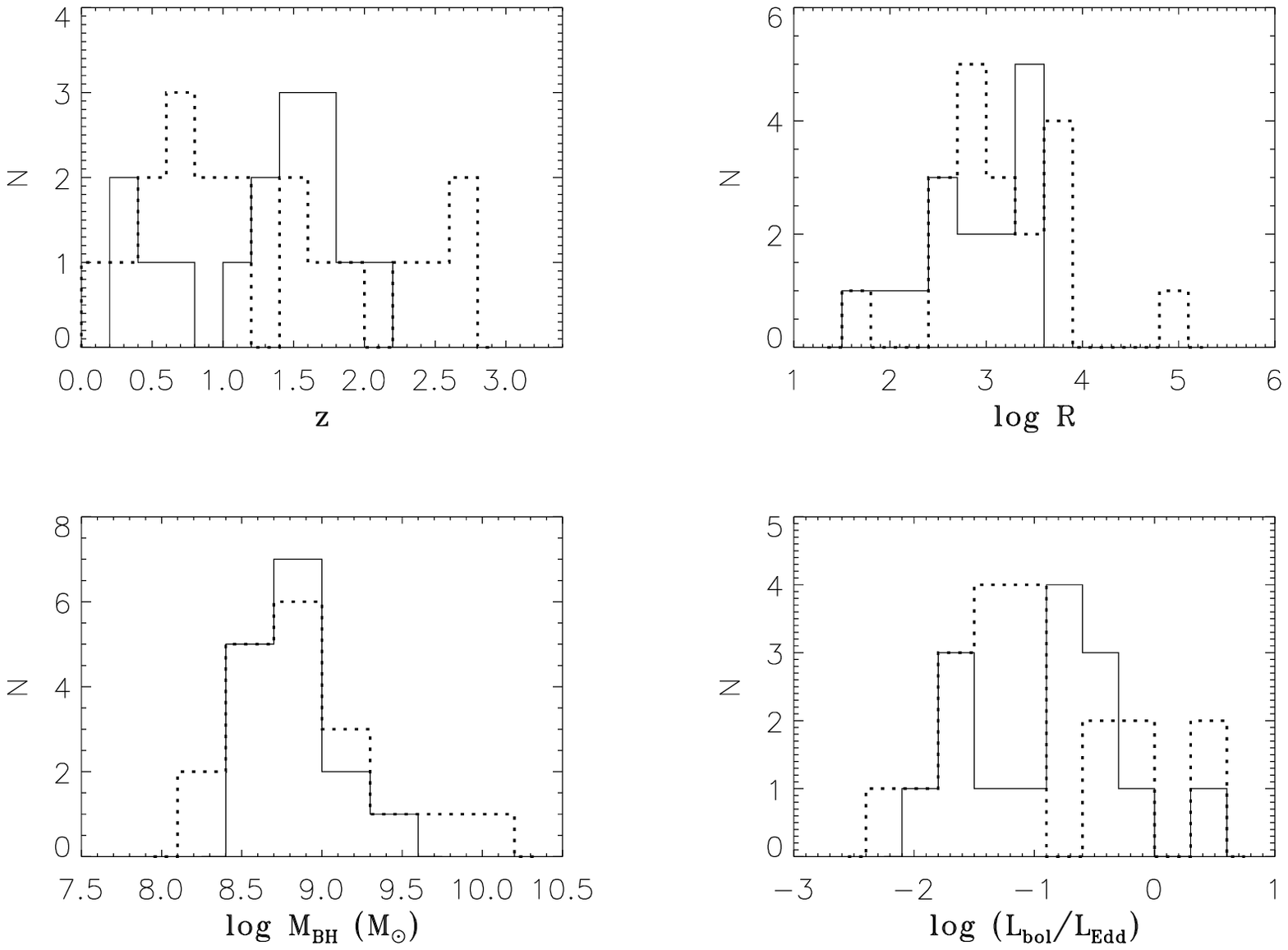}
   \caption{The histogram of sample parameters: redshift (upper left), radio 
loudness (upper right), black hole mass (lower left), and the Eddington 
ratio (lower right). The dotted lines are for 19 SSRQs, and solid lines are for 15 FSRQs.}
              \label{hist}%
    \end{figure}

For completeness, in the following analysis we include five SSRQs 
of Gu \& Ai (2011), and the resulting sample consists of 19 SSRQs. This sample of
19 SSRQs is listed in Table \ref{table_source}, which provides the
source redshift, radio loudness, radio spectral index, black hole
mass, disc bolometric luminosity, and the Eddington ratio. 
The same parameters relative to the 15 FSRQs are listed as a comparison. 
The distribution of redshift, radio loudness, black hole
mass, and the Eddington ratio are shown in Fig. \ref{hist}. We found that SSRQs are 
more extended in the distribution of each parameters than FSRQs.
The source redshift is taken directly from the SDSS DR7 quasar
catalogue, which covers $0.17<z<2.63$ for SSRQs, while $0.34<z<2.00$ for FSRQs. The radio loudness is from
Shen et al. (2011), which ranges from $\rm log~ \it R\rm =1.80$ (SDSS
J013352.66$+$011345.1) to $\rm log ~\it R\rm =5.00$ (SDSS
J000622.60$-$000424.4) for SSRQs, whereas it is 1.55 - 3.54 for FSRQs. However, the radio loudness was calculated
as $R=f_{\rm 6cm}/f_{2500}$, where $f_{\rm 6cm}$ and $f_{2500}$ are
the flux density at rest-frame 6 cm and $2500\AA$, respectively (see
Shen et al. 2011 for more details). Among 15 FSRQs, 4 sources have
an inverted spectral index between 1.4 and 4.85 GHz with $\alpha_{\rm r}<0$.

Black hole masses are estimated from the various empirical relations
in the literature by using the luminosity and FWHM of broad $\rm
H\beta$, Mg II, and C IV lines, i.e., Vestergaard \& Peterson (2006)
for $\rm H\beta$, and Kong et al. (2006) for Mg II and C IV. The
luminosity and FWHM of broad $\rm H\beta$, Mg II, and C IV lines are
adopted from the measurements in Shen et al. (2011). The BLR
luminosity $L_{\rm BLR}$ is derived following Celotti, Padovani \&
Ghisellini (1997) by scaling the strong broad emission lines $\rm
H\beta$, Mg II, and C IV to the quasar template spectrum of Francis
et al. (1991), in which $\rm Ly \alpha$ is used as a flux reference
of 100. By adding the contribution of $\rm H\alpha$ with a value of
77, the total relative BLR flux is 555.77, which consists of $\rm
H\beta$ at 22, Mg II at 34 and C IV at 63 (Francis et al. 1991;
Celotti et al. 1997). From the BLR luminosity, we estimate the disc
bolometric luminosity as $L_{\rm Bol} = 10 L_{\rm BLR}$ (Netzer
1990). It can be seen from Table \ref{table_source} and Fig. \ref{hist} that the black
hole mass of SSRQs ranges from $10^{8.29} M_{\odot}$ to
$10^{10.08} M_{\odot}$ for SSRQs, while $10^{8.44} M_{\odot}$ to
$10^{9.56} M_{\odot}$ for FSRQs. 
The Eddington ratio of SSRQs $\rm log~ \it L_{\rm Bol}/L_{\rm Edd}$
ranges from -2.35 to 0.52, while in FSRQs it ranges from -2.03 to 0.30.

\section{Results}

The SDSS DR7 CAS contains the Stripe82 database, containing all
imaging from SDSS stripe 82 along the celestial equator at the
southern Galactic cap. It includes a total of 303 runs, covering any
given piece of the close to 270 $\rm deg^2$ area approximately 80
times. Only about one-quarter of the stripe 82 scans were obtained
in photometric conditions, the remainder being taken under variable
clouds and often poorer than normal seeing. For the runs that were
non-photometric, an approximate calibration, using the photometric
frames as reference, was derived and made available in the CAS
Stripe82 database. In this work, we directly use the
point-spread-function magnitudes in the CAS Stripe82 database from
the photometric data obtained during the SDSS-I phase from data
release 7 (DR7; Abazajian et al. 2009) and the SN survey during 2005
- 2007. The typical measurement error in magnitude is about 0.03
mag.

Among 19 SSRQs, we select the sources classified as point sources in
all observational runs. Only data with good measurements
(high-quality photometry) are selected following the recommendations
in the SDSS
instructions\footnote{http://www.sdss.org/dr7/products/catalogs/flags.html}.
Moreover, we insist on the $ugriz$ magnitude being brighter than the
magnitude limit in each band, i.e. 22.0, 22.2, 22.2, 21.3, 20.5 at
$u$, $g$, $r$, $i$, $z$, respectively. The data taken at cloudy
conditions are also excluded. We calculate the spectral index $\alpha_{\nu}$ from
the linear fit to the $\rm log~\it f_{\nu} - \rm log ~\nu$ relation
after applying an extinction correction to the $ugriz$ flux density
and taking the flux error into account. In most cases, the linear
fit gave good fits. Each cycle of $ugriz$ photometry was usually
completed within five minutes, i.e. quasi-simultaneously, therefore,
the spectral index calculation will not be seriously influenced by
any source variations.

\subsection{SSRQs}

\subsubsection{Variability}

Among 19 SSRQs, SDSS J013352.66$+$011345.1 was excluded from our
analysis owing to the low quality of the photometric data. All
remaining 18 SSRQs show large amplitude variations with overall
variations in r band $\Delta r=0.22 - 0.92$ mag (see Table
\ref{table_source}). 
In general, the
variations in different bands show similar trends.


The correlation between the spectral index $\alpha_{\nu}$ and PSF
$r$ magnitude was checked for all sources using the Spearman rank
correlation analysis method. We found that 8 of 18 SSRQs show a
significant positive correlation at a confidence level of $>99\%$.
The positive correlation indicates that the
source spectrum becomes flatter when the source is brighter. The result shows that 
a bluer-when-brighter trend is clearly detected in these eight SSRQs. 

\subsubsection{Broad Absorption Line Quasars}

\begin{figure}
   \centering
   \includegraphics[width=0.45\textwidth]{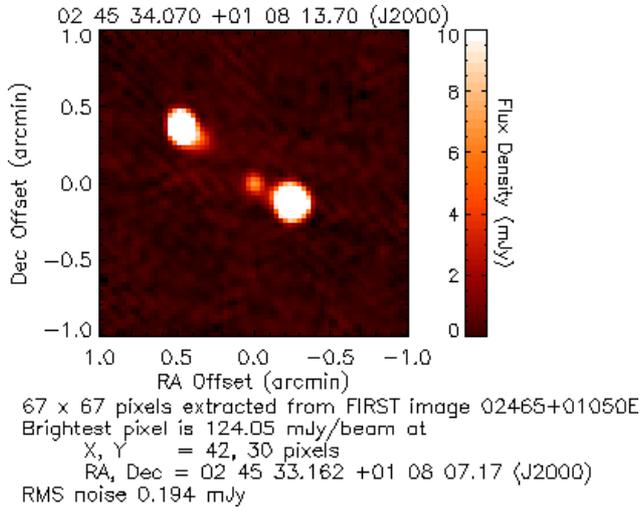}
   \caption{The FIRST radio image of SDSS J024534.07$+$010813.7. }
              \label{j02first}%
    \end{figure}

Among the 18 SSRQs, we found that two sources are defined as broad absorption line 
quasar (BALQ) in the literature.
SDSS J024534.07$+$010813.7 ($z=1.5363$) was listed as a BALQ 
in Gibson et al. (2009) selected from SDSS DR5. 
This source is
also included in the First Bright Quasar Survey with name of 
FBQS J0245+0108 (Becker et al. 2001). It appears as
a single counterpart 
in the NVSS within 1 arcmin of
SDSS position with a flux density of $f_{\rm NVSS}=337.9$ mJy and position
angle of 55.5 degree. 
However, it has three counterparts in the FIRST within 1 arcmin of SDSS positions
 (see Fig. \ref{j02first}). Three counterparts align with a similar position angle with
 that of NVSS counterpart. This difference is simply caused by the different spatial
 resolution of FIRST and NVSS in the case that the radio structure can be resolved in 
 former, however not in the latter. The flux density of the closest FIRST counterpart
is 11.67 mJy, while the integrated flux density of all counterparts is $f_{\rm FIRST}=317.2$ mJy. 
Therefore, the radio emission of this source is highly dominated by the large scale
double radio lobes.
The spectral index calculated for the integrated
FIRST 1.4 GHz and PMN 4.85 GHz flux density of $f_{\rm PMN}=115$ mJy
is $\alpha_{\rm r}=1.05$, while it is $\alpha_{\rm r}=0.82$ when the GB6 4.85 GHz
flux density of $f_{\rm GB6}=85.7$ mJy is used. In contrast, the spectral index 
will be highly inverted if only the flux density of the closest FIRST counterpart is used.

\begin{figure}
   \centering
   \includegraphics[width=0.45\textwidth]{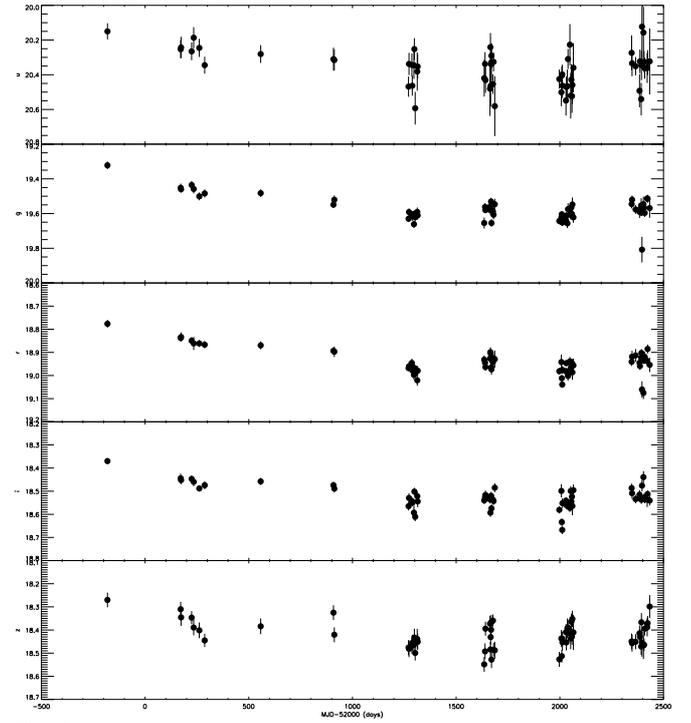}
   \caption{The $ugriz$ band light curve of SDSS J024534.07$+$010813.7 (from top to bottom). }
              \label{j02lc}%
    \end{figure}

The light curves of SDSS J024534.07$+$010813.7 in u, g, r, i, and z
bands are shown in Fig. \ref{j02lc}. The variations exhibit similar
trends in all bands, over seven years from 2000 to 2007. The overall
variation in r band is 0.29 mag. In the observational sessions, the source 
became gradually fainter from 2000 to 2004, and then stayed quite stable
from 2004 to 2007.

\begin{figure}
   \centering
   \includegraphics[width=0.45\textwidth]{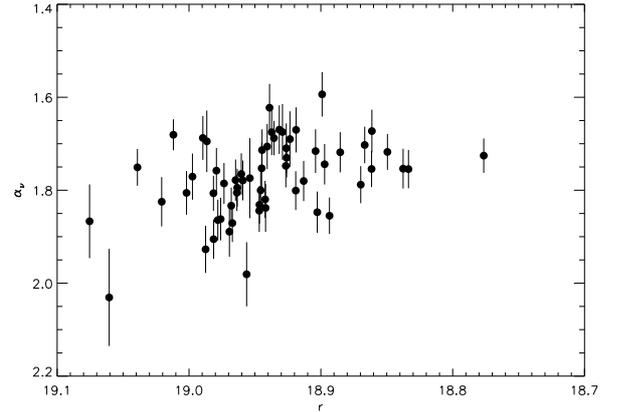}
   \caption{The relationship between the spectral index and the PSF magnitude
   at r band for SDSS J024534.07+010813.7. A significant positive correlation is
   present, which implies a bluer-when-brighter trend.}
              \label{j02ra}%
    \end{figure}

In SDSS J024534.07$+$010813.7, a significant positive correlation present
 between the spectral index $\alpha_{\nu}$
and $r$ magnitude, which is shown in Fig. \ref{j02ra}. The Spearman
correlation analysis shows a significant positive correlation with a
correlation coefficient of $r_{\rm s}=0.435$ at confidence level of
$>99.9\%$. The positive correlation shows that the
source spectrum becomes flatter when the source is brighter, i.e.
that there is a bluer-when-brighter trend. With the source redshift
$z=1.5363$, SDSS $ugriz$ wavebands correspond to the wavelength
range of $1400 - 3521 \AA$ in the source rest-frame. For a                          
sample of 17 radio-quiet AGNs, Shang et al. (2005) show that the
spectral break occurs at around $1100~ \AA$ for most objects, which
is thought to be closely related to the big blue bump. If this
spectral break also exists in SDSS J024534.07$+$010813.7, we would
expect to observe the rising part of accretion disk thermal emission
when it dominates over the nonthermal emission. However, the spectral index
in the observational sessions are in the range of 1.59 to 2.03. This implies a
falling SEDs in the rest frame $1400 - 3521 \AA$. This can be likely 
caused by the internal extinction in quasar itself, for example, from the 
outflowing absorption gas.

We found another BALQ SDSS J021728.62$-$005227.2 ($z=2.4621$) in SSRQs of Gu \& Ai (2011).  
It is defined as BALQ in Trump et al. (2006) selected from SDSS DR3. It is defined
as SSRQs in Gu \& Ai (2011) with radio spectral index of $\alpha_{\rm r}=0.75$. Its overall 
variations amplitude in r band is $\Delta r=0.35$. The significant positive correlation between
the r magnitude and the optical spectral index shows a bluer-when-brighter trend in this
source. However, unlike SDSS J024534.07$+$010813.7, the spectral index varies from 0.45 to 1.28, 
with most time staying at $<1.0$. The internal extinction from the outflowing absorption
gas may not be severe in this source.

\subsection{FSRQs}

All 15 FSRQs show large amplitude variations with overall
variations in r band $\Delta r=0.18 - 0.97$ (see Table
\ref{table_source}). The correlation between the spectral index $\alpha_{\nu}$ and PSF
$r$ magnitude was checked for all sources using the Spearman rank
correlation analysis method. We found that 11 of 15 FSRQs show a
significant positive correlation at a confidence level of $>99\%$, i.e. 
a bluer-when-brighter trend.


\subsection{BLR geometry}

\begin{figure}
   \centering
   \includegraphics[width=0.45\textwidth]{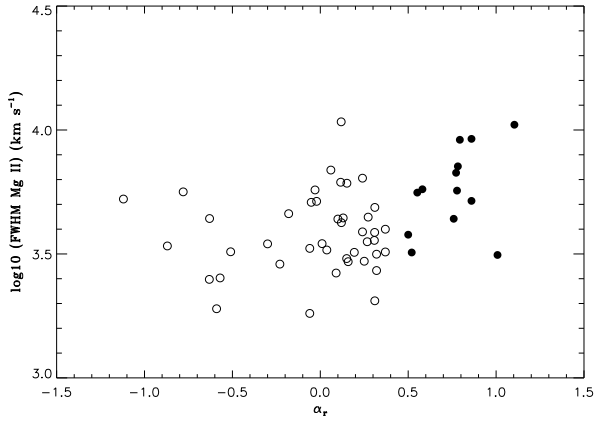}
   \caption{The line width of broad Mg II line versus radio spectral index. The solid circles represent
   SSRQs, while the open circles are for FSRQs.}
              \label{alfwhm}%
    \end{figure}

Assuming the radio spectral index as an indicator of orientation, the BLR
geometry can be investigated when line width of broad emission lines are included 
(Fine et al. 2011). To this end, we combine all sources in this work and those 
FSRQs in Gu \& Ai (2011), resulting a sample of 62 quasars, which includes 
18 SSRQs and 44 FSRQs.
The relationship between the radio spectral index and FWHM of Mg II
line is shown in Fig. \ref{alfwhm} for a subsample of 55 sources (42 FSRQs and 13 SSRQs) 
with available Mg II line measurements. There is a trend of broader Mg
II lines with steeper spectral index. The Spearman rank correlation
analysis shows a positive correlation with correlation coefficient
of $r_{\rm s}=0.298$ at confidence level of 97.3\%. The mean value
of Mg II FWHM is 4075 $\rm km~ s^{-1}$ for FSRQs, while it is 6110
$\rm km~ s^{-1}$ for SSRQs. The median value of Mg II FWHM is 3586
$\rm km~ s^{-1}$ and 5694 $\rm km~ s^{-1}$ for FSRQs and SSRQs, respectively.
Although our sample is much smaller than theirs, the results are consistent 
with that of Fine et al. (2011), which used the radio spectral index between 
Westerbork Northern Sky Survey 330 MHz and NVSS 1.4 GHz. The result implies a
disk-like BLR geometry may present in these quasars.

\section{Discussions}

\subsection{SSRQs and FSRQs}

\begin{figure}
   \centering
   \includegraphics[width=0.45\textwidth]{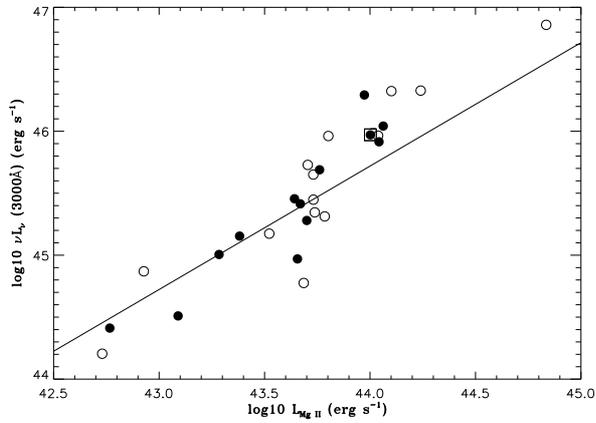}
   \caption{The plot of broad Mg II line luminosity and continuum luminosity
   at 3000 $\AA$ for 13 SSRQs and 14 FSRQs. The symbols are same as in Fig. 4. The square is for
   BALQ SDSS J024534.07$+$010813.7. 
   The solid line is the OLS bisector linear fit to radio-quiet
   AGNs in Kong et al. (2006), $\lambda L_{\rm \lambda~ 3000 \AA}=78.5~ L^{0.996}_{\rm Mg~ II}$.} 
              \label{mgcon}%
    \end{figure}

While FSRQs are usually associated with core-dominated radio
quasars, SSRQs are generally related to lobe-dominated ones, usually
with two large-scale optically thin radio lobes. Usually, the
beaming effect is not severe in SSRQs because of the relatively
large viewing angle. Therefore, the optical continuum of SSRQs could
be dominated by the thermal emission. In Fig. \ref{mgcon}, most SSRQs
with available Mg II measurements lie close to or  below the
solid line for radio quiet AGNs. The average value of the ratio
of the luminosity at 3000 $\AA$ to that of broad Mg II line is
$<\rm log~ \it (\nu L_{\nu}(\rm 3000\AA)/\it L_{\rm Mg II})>\rm =1.78\pm0.26$, which is in good agreement 
with the linear relation for radio quiet AGNs (see Fig. \ref{mgcon}). This likely implies that the
thermal emission from accretion disk is indeed the dominant one in the
optical continuum at the epoch of SDSS spectra for most SSRQs, if not all. 
For comparison, 14 FSRQs with Mg II measurements are also plotted in Fig. \ref{mgcon}. 
We found that the distribution of FSRQs are generally similar to that of SSRQs, but with larger scatter. 
The average value of the ratio
of the luminosity at 3000 $\AA$ to that of broad Mg II line is
$<\rm log~ \it (\nu L_{\nu}(\rm 3000\AA)/\it L_{\rm Mg II})>\rm =1.81\pm0.31$, which is slightly larger
than that of SSRQs. This may imply that the optical nonthermal jet emission may not be dominant 
in all FSRQs, as found by Chen et al. (2009). 

\begin{figure}
   \centering
   \includegraphics[width=0.45\textwidth]{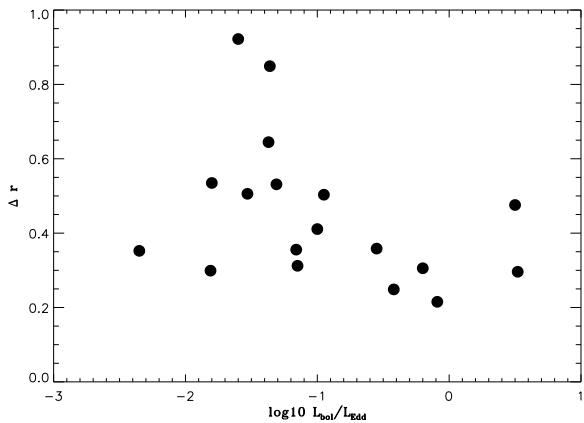}
   \caption{The Eddington ratio versus the variability at r band $\Delta r$ for SSRQs.}
              \label{ledr}%
    \end{figure}

Interestingly, eight
of 18 SSRQs display a BWB trend (see Table \ref{table_source}).
The BWB trend has also been found in radio quiet AGNs, and the
variability was found to be anti-correlated with the Eddington ratio
(Ai et al. 2010). In radio quiet AGNs, the optical variability could
be due to the variation in the accretion process, for example, the
variation in the accretion rate (e.g. Li \& Cao 2008), since the jet
is either weak or absent. These explanations could also be used to
explain the variability of SSRQs if the optical emission is indeed
mainly from the accretion disk. 
To further investigate the mechanism of variability in SSRQs, 
the relationship between the variability $\Delta r$ and 
the Eddington ratio is plotted in Fig. \ref{ledr}. We found an anti-correlation 
with a Spearman correlation coefficient of $r_{\rm s}=-0.490$ at 
confidence level of $96.2\%$. 
This similar anti-correlation to radio quiet AGNs, strongly implies that the 
mechanisms of optical variability may be similar in the two populations. The thermal emission 
from the accretion disk may be responsible for the variability of SSRQs. 
In contrast, there is no correlation for FSRQs. This implies that 
the mechanism of variability in FSRQs may be different from that of SSRQs.
However, that the variation in SSRQs is
caused by jet nonthermal emission cannot be completely excluded in
some cases because the variability amplitudes are generally larger
that the typical values for radio quiet AGNs, 0.05 - 0.3 mag (e.g. Ai et al. 2010), 
 (see Table \ref{table_source}). In extreme cases, the overall amplitude is
$\Delta r=0.92$ mag in SDSS J013514.39$-$000703.8. It may be more
likely that both thermal and nonthermal emission contribute to the
variability. Further multi-waveband monitoring is needed to help
resolve these uncertainties, especially the spectroscopic
monitoring.

In terms of optical variability on month-to-year time-scales, no obvious difference was found in 
the behavior of radio quiet quasars and lobe-dominated quasars (LDQs)  (Stalin et al. 2004). 
Moreover, there is no significant difference in either the amplitude or duty
cycle of intranight optical variability (INOV) between these two classes of non-blazar AGNs. 
However, their sample is too small, and the time spanned by observations is too short. 
Nevertheless, the authors infer that the radio loudness of a quasar alone is not a sufficient condition for a pronounced
INOV. While the observational data cannot exclude accretion disc flares as the source of the
intranight optical variability in radio quiet quasars and LDQs, it does not preclude a substantial
contribution from blazar-like relativistically beamed emission. As suggested by Stalin et 
al. (2005), the radio quiet quasars may also eject relativistic jets. However, their jets are 
probably quenched while crossing the innermost micro-arcsecond scale, possibly
through heavy inverse Compton losses in the vicinity of the central engine (e.g. Gopal-Krishna et al. 2003).
However, the analysis of spectra of a set of radio quiet quasars that had already been 
searched for microvariability showed that the jet-based scenario is unlikely (Chand, Wiita \& Gupta 2010). 

\begin{figure}
   \centering
   \includegraphics[width=0.45\textwidth]{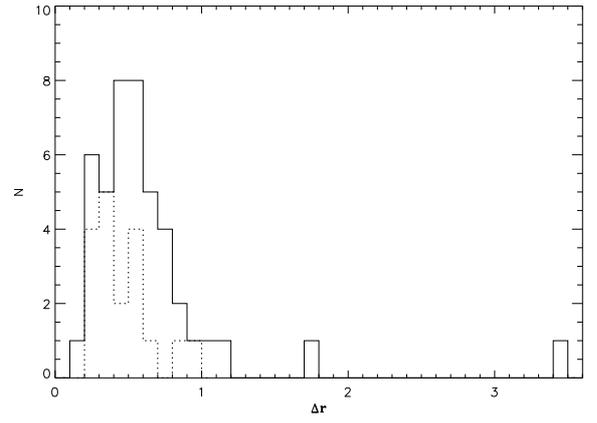}
   \caption{The histogram of $\Delta r$ for SSRQs and FSRQs in Gu \& Ai (2011) and in this work. 
   The dotted lines are for 18 SSRQs, and solid lines are for 44 FSRQs.}
              \label{drhist}%
    \end{figure}

In order to compare systematically the variations of SSRQs with FSRQs, 
we combine the FSRQs of Gu \& Ai (2011) with sources in this work. 
The combined sample consists of 62 quasars, of which 44 are FSRQs, and
18 are SSRQs. We show the histogram of variation amplitude
in r band $\Delta r$ for both SSRQs and FSRQs in Fig. \ref{drhist}. 
It can be seen that $\Delta r$ of FSRQs are systematically larger than that of
SSRQs, with median value of 0.52 mag and 0.41 mag for FSRQs and SSRQs, respectively.
Apparently, FSRQs cover wider range in $\Delta r$. Four FSRQs exhibit $\Delta r>1.0$ mag with an extreme 
case of $\Delta r=3.46$ (see Gu \& Ai 2011), while none in SSRQs. This 
implies that the variations in FSRQs are systematically more violent than SSRQs.
This is most likely due to the relativistic jet, which is viewed at smaller viewing
angle in FSRQs. The fact that the typical variations amplitude of SSRQs is between
the values of radio quiet AGNs and FSRQs is qualitatively consistent with the source
nature of SSRQs, which are radio-loud sources with prominent powerful relativistic jets, 
however viewed at larger viewing angle resulting in a dominated thermal emission
at optical wavebands. In terms of the spectral variations, 
eight of 18 SSRQs show BWB trend at confidence level
of $>99\%$, i.e. at probability of $\sim 44\%$. In contrast, 25 ($\sim57\%$) of the
44 FSRQs show BWB trend at confidence level of $>99\%$, while only one source shows a RWB trend
at confidence level of $>99\%$. Therefore, it seems that BWB trend are quite 
common both in FSRQs and in SSRQs, although the mechanisms can be different, for
example, shock-in-jet and accretion rate change are mainly responsible for FSRQs and
SSRQs, respectively.

In the unification scheme of AGNs, the radio galaxies are generally unified with radio quasars 
(Antonucci 1993), with radio galaxies being viewed at large viewing angle. The jet emission in radio galaxies
are not severely boosted, then the dominated optical emission is expected to be thermal emission related to
the accretion disk. Indeed, it is argued that the optical and X-ray emission arise from the accretion disk - 
corona system in two radio galaxies 3C 120 and 3C 111 by using  extensive multi-frequency monitoring data (Chatterjee et al. 2009, 2011). 
In both sources, the low optical polarization supports the thermal origin of the optical emission. 
However, in 3C 111 the detection of polarization percentages of 3\% or even higher suggests that during 
particular activity phases an additional contribution from non-thermal emission from the jet may be 
present (Jorstad et al. 2007; Chatterjee et al. 2011). Interestingly, 
signiÞcant dips in the X-ray light curve are followed by ejections of bright superluminal knots in the VLBA images 
for both sources. The authors argued that the radiative state of accretion disk plus corona system, where the X-rays are produced, has a direct 
effect on the events in the jet, where the radio emission originates.  

Similar to many FSRQs in Gu \& Ai (2011) having a spectral index 
$\alpha_{\nu}<1.0$, almost all FSRQs in this work show $\alpha_{\nu}<1.0$, implying
a rising SED in optical regions. One possibility is that the optical emission
is mainly from thermal accretion disc, although these sources are defined as
FSRQs. Alternatively, they can be high-synchrotron-peaked FSRQs, since
the synchrotron peak frequency is higher than the rest-frame frequency of SDSS
$ugriz$ wavebands. This however needs further investigations.

\subsection{BALQs}

The nature of BAL quasars is still unclear. It was argued that the outflowing BAL wind
is preferentially equatorial, and only in those objects with almost edge-on accretion disk to 
the line of sight, can be observed as BAL (Cohen et al. 1995; Goodrich \& Miller 1995).  
However, the VLBI radio images of BALQs argued against this orientation scenario, in 
terms of the radio structures and both the steep and flat radio spectrum (e.g Jiang \& Wang 2003; Liu et al. 2008; Doi et al. 2009).
Fine et al. (2011)  found no evidence that BAL QSOs have a different spectral index 
distribution to non-BALs although only 25 obvious BALs are considered in their sample.  
Their BALQs both have steep and flat radio spectrum computed using the Westerbork Northern Sky Survey (WENSS)
at 330 MHz and the NVSS at 1.4 GHz. In general, the radio-loud BALQs tend to be compact in the radio, similar to 
Gigahertz-Peaked Spectrum (GPS) and Compact Steep Spectrum (CSS) sources, which are thought to be
at the young stage of large scale radio sources (Becker et al. 2000; Montenegro-Montes et al. 2009). 
It has been proposed that the BALs are not closely related with the inclinations, and may be 
associated with a relatively short-lived evolutionary phase with a large BAL wind covering fraction 
(e.g. Briggs et al. 1984; Gregg et al. 2000). We found that the two radio-loud steep spectrum BALQs in this paper have different radio structures. 
SDSS J024534.07$+$010813.7 has prominent extended double radio lobes, of which the emission dominates over
that of radio core (see Fig. \ref{j02first}). The source is too extended 
($\sim 52$ arcsec, i.e. $\sim 440$ kpc) to be considered as a CSS/GPS. This source has not been imaged with VLBI.
In contrast, SDSS J021728.62$-$005227.2 is compact in FIRST image with only single counterpart
found. It has been imaged with VLBI by Doi et al. (2009). However, only a compact component was found 
with flux density at 8.4 GHz $f_{\rm 8.4GHz}=43$ mJy at scale of $<1$ kpc. In combination with the 
FIRST flux density, the radio spectrum is steep $\alpha_{\rm r}=0.9$, which makes this source resemble to CSS, thus likely
being a young radio source. More radio observations at higher spatial resolution would be helpful to
better understand the nature of this source.

While there are many investigations on the variability of broad absorption 
line troughs (e.g. Barlow 1993; Lundgren et al. 2007;  Gibson et al. 2008, 2010; Capellupo et al. 2011), 
not much works have been done on the optical continuum variability for BALQs.
On timescales of the order of years, the optical continuum variability of BALQs was found
to differ markedly from that of non-BAL radio quiet quasars (Turnshek 1988).
In short timescales, intranight optical variations of $\sim5\%-9\%$ were detected on timescales of 
$\sim 1$ hr in two BALQs (Anupama \& Chokshi 1998). 
The recent search for optical microvariability in a sample of 19 
radio quiet BALQs argued that radio quiet BALQs do not appear to be a special case of the radio
quiet quasars in terms of their similar duty cycle (Joshi et al. 2011). 
However, all these observations focused only on the radio quiet BALQs. 
Therefore, for the first time, we present here the optical continuum variability 
for radio loud BALQs, although only two sources are found in our sample.
In both BALQs, the C IV BAL troughs fall outside the $ugriz$ bands. Together with the similar trends 
of variability in different wavebands rule out the possibility of
the observed light-curve variations being caused by variations in the strength of the BAL 
features. The variation amplitude in the r band of two sources is not atypical in overall sample
of SSRQs, although they are at the low end of distribution in Fig. \ref{drhist}. This 
implies that the mechanism of optical continuum variability of radio-loud BALQs may likely 
be similar to that of more general SSRQs. Interestingly, both BALQs show BWB trend. 
The percentage of BWB trend is higher than that of SSRQs. However, the small sample size
precludes to draw any firm conclusions. Certainly, a larger sample is needed for further investigations. 
The spectral index of two BALQs in this work are steeper than the typical value of other SSRQs, which implies
more reddened in BALQs than non-BALQs. This is consistent with previous results (Trump et al. 
2006; Gibson et al. 2009).

\section{Summary}

We have constructed a sample of 19 SSRQs and 15 FSRQs in the SDSS Stripe 82
region. The variability and the relationship between the spectral
index and brightness were investigated for 18 SSRQs and 15 FSRQs. We found that
all SSRQs show large-amplitude overall variations, e.g. from 0.22 to
0.92 mag in r band. We found a bluer-when-brighter trend in 8
of the 18 SSRQs studied here. 
In contrast, the variability amplitude of 15 FSRQs in r band ranges from 0.18 to 0.97 mag, 
and 11 of 15 sources show a BWB trend.
We found an anti-correlation between the Eddington ratio and the variability amplitude 
in r band for SSRQs, which is similar to that in radio quiet AGNs. This implies that the thermal 
emission from accretion disk may be responsible for the variability in SSRQs, although the jet nonthermal 
emission cannot be excluded. In contrast, there is no correlation in FSRQs, which implies that the mechanisms of 
variability in FSRQs may be different from that in SSRQs. The optical continuum variability of radio loud
BALQs is investigated for the first time here on two sources with steep
radio spectrum. Both radio loud BALQs show variations with amplitude of about 0.3 mag
at r band. Their spectral variability all show a bluer-when-brighter trend.  In combination with 29 FSRQs also selected
from Stripe 82 region in Gu \& Ai (2011), we found a trend of a broader line width of broad Mg II line with steeper 
radio spectral index. This implies a disc-like BLR geometry may present in these quasars. 
In these 62 quasars, we found a $\sim57\%$ source percentage showing BWB trend in FSRQs, whereas it 
is $\sim44\%$ in SSRQs.

%

\begin{acknowledgements}
We thank the anonymous referee for constructive comments that 
improved the manuscript.
MFG thanks X. Cao, S. Li, A. Gupta, J. Wu and Z. Chen for useful
discussions. This work is supported by the National Science
Foundation of China (grants 10703009, 10821302, 10833002, 10978009,
11033007 and 11073039), and by the 973 Program (No. 2009CB824800).
Funding for the SDSS and SDSS-II was provided by the Alfred P. Sloan
Foundation, the Participating Institutions, the National Science
Foundation, the U.S. Department of Energy, the National Aeronautics
and Space Administration, the Japanese Monbukagakusho, the Max
Planck Society, and the Higher Education Funding Council for
England. The SDSS Web site is http://www.sdss.org/.
\end{acknowledgements}

\clearpage

\newpage


\clearpage

\newpage

\begin{table}
\caption{\label{table_source}Source list: Col. 1 - SDSS source name;
Col. 2 - redshift; Col. 3 - radio loudness; Col. 4 - the spectral
index between 1.4 and 4.85 GHz, of which GB6 flux density is used
for those labeled with $^{a}$, otherwise PMN one is used; Col. 5 -
black hole mass in unit of solar mass; Col. 6 - disc bolometric
luminosity in unit of $\rm erg~s^{-1}$; Col. 7 - the Eddington ratio
$l=L_{\rm BOL}/L_{\rm EDD}$; Col. 8 - overall variation in r band;
Col. 9 - the range of variation in the spectral index
$\alpha_{\nu}$; Cols. (10-11) - the Spearman correlation coefficient
and probability level,
respectively.
} \centering
\begin{tabular}{lcclccrcrrl}
\hline\hline
SDSS source   &    $z$   & log 10 (R) & $\alpha_{\rm r}$ & log 10 ($M_{\rm BH}$) & log 10 ($L_{\rm BOL}$) & log 10 ($l$) & $\Delta~ r$ & $\Delta~ \alpha_{\nu}$ & $r_{\rm s}$ & prob. \\
 & & & & ($\rm M_{\odot}$) & ($\rm erg~s^{-1}$) &  & (mag) & & \\ 
(1)&(2)&(3)&(4)&(5)&(6)&(7)&(8)&(9)&(10)&(11)\\
\hline
&&&& SSRQs &&&&&& \\
\hline
J000622.60$-$000424.4  &   1.0377   &   5.00 &   0.77        &  8.92   &    45.87   &   -1.16  &   0.35  &  0.28,1.00  &   0.146  &   0.24     \\
J005905.51$+$000651.6  &   0.7189   &   3.71 &   0.50        &  8.96   &    45.92   &   -1.15  &   0.31  & -0.16,0.49  &   0.315  &   1.32e-02 \\
J013352.66$+$011345.1  &   0.3081   &   1.80 &   0.54$^{a}$  &  8.34   &    45.10   &   -1.35  &   ...   &  ...        &   ...    &    ...     \\
J013514.39$-$000703.8  &   0.6712   &   3.73 &   0.75        &  8.41   &    44.92   &   -1.60  &   0.92  &  0.36,1.30  &   0.686  &   7.08e-10 \\
J015509.00$+$011522.5  &   1.5480   &   2.99 &   0.51$^{a}$  &  8.64   &    46.55   &   -0.20  &   0.30  &  0.46,0.84  &   0.391  &   7.44e-04 \\
J021225.56$+$010056.1  &   0.5128   &   2.56 &   0.55$^{a}$  &  8.77   &    45.51   &   -1.37  &   0.64  & -0.08,0.28  &   0.275  &   0.03     \\
J023313.81$-$001215.4  &   0.8072   &   2.93 &   0.78        &  8.41   &    45.57   &   -0.95  &   0.50  &  0.06,0.54  &   0.282  &   1.10e-02 \\
J024534.07$+$010813.7  &   1.5363   &   3.16 &   1.10$^{a}$  &  9.65   &    45.95   &   -1.81  &   0.29  &  1.59,2.03  &   0.435  &   5.08e-04 \\
J213004.75$-$010244.4  &   0.7040   &   3.25 &   0.79        & 10.08   &    45.85   &   -2.35  &   0.35  &  0.45,0.97  &   0.052  &   0.60     \\
J213513.10$-$005243.8  &   1.6548   &   2.74 &   1.00        &  8.58   &    46.60   &   -0.09  &   0.22  &  0.40,0.77  &   0.279  &   0.02     \\
J221409.96$+$005227.0  &   0.9078   &   2.87 &   0.77$^{a}$  &  9.05   &    45.86   &   -1.31  &   0.53  & -0.09,0.43  &   0.405  &   1.94e-03 \\
J231607.25$+$010012.9  &   2.6291   &   2.56 &   0.82$^{a}$  &  9.15   &    46.84   &   -0.42  &   0.24  &  0.38,0.89  &   0.445  &   8.31e-04 \\
J233624.04$+$000246.0  &   1.0949   &   2.67 &   0.58        &  9.16   &    46.28   &   -1.00  &   0.41  &  0.12,0.61  &   0.787  &   1.37e-10 \\
J235156.12$-$010913.3  &   0.1739   &   2.81 &   0.68        &  8.90   &    45.48   &   -1.53  &   0.50  & -0.13,0.31  &   0.090  &   0.54     \\
\hline
&&& SSRQs & from & Gu \& Ai (2011) &&&&&\\
\hline
J012401.76$+$003500.9  &  1.8516    &  3.85  &    0.86$^{a}$ &   9.36   &    46.11   &   -1.36  &   0.84   &  -0.52,0.61   &    0.687   &  6.8e-07  \\
J012517.14$-$001828.9  &  2.2780    &  3.44  &    0.59       &   8.50   &    47.13   &    0.52  &   0.29   &  0.22,0.87    &   -0.202   &  0.11      \\
J015832.51$-$004238.2  &  2.6071    &  3.58  &    0.87       &   8.29   &    46.90   &    0.50  &   0.47   &  0.44,1.19    &   -0.207   &  0.28      \\
J021728.62$-$005227.2  &  2.4621    &  3.29  &    0.75       &   8.84   &    46.40   &   -0.55  &   0.35   &  0.45,1.28    &    0.488   &  7.7e-06  \\
J022508.07$+$001707.2  &  0.5270    &  3.81  &    0.86$^{a}$ &   8.88   &    45.19   &   -1.80  &   0.53   &  0.25,0.68    &    0.470   &  0.03  \\
\hline
&&&& FSRQs &&&&&& \\
\hline
J000111.19$-$002011.5  &   0.5179   &   2.58 &   0.15        &  8.57   &    45.16   &   -1.52  &   0.36  &  0.64,1.04  &   0.426  &   5.44e-03 \\
J010033.50$+$002200.1  &   0.7534   &   1.94 &   0.11$^{a}$  &  9.21   &    45.93   &   -1.40  &   0.55  & -0.33,0.26  &   0.615  &   2.39e-08 \\
J010826.84$-$003724.2  &   1.3724   &   3.31 &   0.26        &  8.76   &    46.31   &   -0.56  &   0.65  &  0.25,0.80  &   0.330  &   4.00e-03 \\
J015105.80$-$003426.4  &   0.3352   &   1.55 &  -0.24        &  8.73   &    44.82   &   -2.03  &   0.55  &  0.05,0.95  &   0.807  &   2.39e-12 \\
J020234.32$+$000301.7  &   0.3664   &   3.26 &   0.31        &  8.47   &    44.81   &   -1.77  &   0.18  &  0.19,0.58  &  -0.274  &   0.36     \\
J020326.98$+$003744.3  &   1.5840   &   3.43 &   0.30$^{a}$  &  8.44   &    46.31   &   -0.24  &   0.80  &  0.24,0.89  &   0.678  &   2.71e-05 \\
J021612.20$-$010518.9  &   1.4931   &   2.61 &   0.03        &  8.77   &    46.45   &   -0.43  &   0.27  &  0.37,0.73  &   0.263  &   0.04     \\
J024854.81$+$001053.8  &   1.1457   &   2.92 &   0.30        &  8.65   &    46.00   &   -0.77  &   0.97  & -0.31,0.56  &   0.723  &   7.06e-12 \\
J025928.51$-$001959.9  &   2.0001   &   2.54 &  -0.63        &  8.87   &    47.29   &    0.30  &   0.30  &  0.08,0.66  &   0.323  &   1.24e-02 \\
J031318.66$+$003623.9  &   1.2561   &   2.35 &   0.11$^{a}$  &  9.56   &    46.02   &   -1.65  &   0.32  &  0.52,0.98  &   0.353  &   4.51e-03 \\
J220719.77$+$004157.3  &   1.8926   &   3.36 &  -0.05        &  8.85   &    45.90   &   -1.06  &   0.59  &  0.05,0.58  &  -0.171  &   0.18     \\
J222729.05$+$000521.9  &   1.5133   &   3.02 &   0.15        &  9.19   &    46.52   &   -0.79  &   0.70  &  0.46,1.36  &   0.722  &   3.18e-10 \\
J233200.00$+$011510.9  &   1.6395   &   3.45 &   0.27$^{a}$  &  8.75   &    46.01   &   -0.86  &   0.44  &  0.15,0.87  &   0.647  &   8.97e-08 \\
J234624.56$+$001914.2  &   1.7775   &   2.84 &  -0.02$^{a}$  &  8.87   &    46.39   &   -0.60  &   0.79  &  0.09,0.72  &   0.702  &   9.77e-11 \\
J234830.98$+$011037.6  &   1.7064   &   3.54 &   0.19$^{a}$  &  8.44   &    45.87   &   -0.68  &   0.60  &  0.17,1.08  &   0.746  &   1.67e-07 \\
\hline
\end{tabular}
\end{table}


\begin{thebibliography}{}

\bibitem[Abazajian et al. (2009)]{aba09} Abazajian, K. N., Adelman-McCarthy, J. K., Ag\"{u}eros, M. A., et al. 2009, ApJS, 182, 543
\bibitem[Ai et al. (2010a)]{ai10a} Ai, Y. L., Yuan, W., Zhou, H. Y., et al. 2010, ApJ, 716, L31
\bibitem[Antonucci(1993)]{1993ARA&A..31..473A} Antonucci, R. 1993, ARA\&A, 31, 473 
\bibitem[Anupama \& Chokshi(1998)]{anu98} Anupama, G. C., \& Chokshi, A. 1998, ApJ, 494, L147
\bibitem[Barlow(1993)]{bar93} Barlow, T. A., 1993, PhD thesis, Univ. California, San Diego 
\bibitem[Becker, White \& Helfand(1995)]{bec95} Becker, R. H., White, R. L., \& Helfand, D. J. 1995, ApJ, 450, 559
\bibitem[Becker et al.(2000)]{bec00} Becker, R. H., White, R. L., Gregg, M. D., et al. 2000, ApJ, 538, 72 
\bibitem[Becker et al. (2001)]{bec01} Becker, R. H., White, R. L., Gregg, M. D., et al. 2001, ApJS, 135, 227
\bibitem[Briggs et al.(1984)]{bri84} Briggs, F. H., Turnshek, D. A., \& Wolfe, A. M. 1984, ApJ, 287, 549
\bibitem[Capellupo et al.(2011)]{cap11} Capellupo, D. M., Hamann, F., Shields, J. C., Rodr\'{i}guez Hidalgo, P., \& Barlow, T. A. 2011, MNRAS, 413, 908
\bibitem[Celotti, Padovani \& Ghisellini(1997)]{cel97} Celotti, A., Padovani, P., \& Ghisellini, G., 1997, MNRAS, 286, 415
\bibitem[Chand et al.(2010)]{cha10} Chand, H., Wiita, P. J., \& Gupta, A. C. 2010, MNRAS, 402, 1059
\bibitem[Chatterjee et al.(2009)]{cha09} Chatterjee, R., Marscher, A. P., Jorstad, S. G., et al. 2009, ApJ, 704, 1689
\bibitem[Chatterjee et al.(2011)]{cha11} Chatterjee, R., Marscher, A. P., Jorstad, S. G., et al. 2011, ApJ, 734, 43 
\bibitem[Chen et al.(2009a)]{che09a} Chen, Z. Y., Gu, M. F., \& Cao, X. 2009, MNRAS, 397, 1713
\bibitem[Clements \& Carini(2001)]{cle01}Clements, S. D., \& Carini, M. T. 2001, AJ, 121, 90
\bibitem[Cohen et al.(1995)]{coh95} Cohen, M. H., Ogle, P. M., Tran, H. D., et al. 1995, ApJ, 448, L77
\bibitem[D'Ammando et al.(2011)]{2011A&A...529A.145D} D'Ammando, F., Raiteri, C. M., Villata, M., et al. 2011, A\&A, 529, A145 
\bibitem[Doi et al.(2003)]{doi03} Doi, A., Kawaguchi, N., Kono, Y., et al. 2009, PASJ, 61, 1389
\bibitem[Fan et al.(1998)]{fan98} Fan, J. H., Xie, G. Z., Pecontal, E., Pecontal, A., \& Copin, Y. 1998, ApJ, 507, 173
\bibitem[Fine et al.(2011)]{fin11} Fine, S., Jarvis, M. J., \& Mauch, T. 2011, MNRAS, 412, 213
\bibitem[Francis et al.(1991)]{Francis91} Francis, P.~J., Hewett, P.~C., Foltz, C.~B., et al. 1991, ApJ, 373, 465
\bibitem[Ghisellini et al.(1997)]{ghi97}Ghisellini, G., Villata, M., Raiteri, C. M., et al. 1997, A\&A, 327, 61
\bibitem[Ghosh et al.(2000)]{gho00}Ghosh, K. K., Ramsey, B. D., Sadun, A. C., \& Soundararajaperumal, S. 2000, ApJS, 127, 11
\bibitem[Gibson et al.(2008)]{gib08} Gibson, R. R., Brandt, W. N., Schneider, D. P., \& Gallagher, S. C. 2008, ApJ, 675, 985
\bibitem[Gibson et al.(2009)]{gib09} Gibson, R. R., Jiang, L., Brandt, W. N., et al. 2009, ApJ, 692, 758
\bibitem[Gibson et al.(2010)]{gib10} Gibson, R. R., Brandt, W. N., Gallagher, S. C., Hewett, P. C., \& Schneider, D. P. 2010, ApJ, 713, 220
\bibitem[Goodrich \& Miller(1995)]{goo95} Goodrich, R. W., \& Miller, J. S., 1995, ApJ, 448, L73
\bibitem[Gopal-Krishna et al.(1995)]{gop95} Gopal-Krishna, Sagar, R., \& Wiita, P. J. 1995, MNRAS, 274, 701
\bibitem[Gopal-Krishna et al.(2003)]{gop03} Gopal-Krishna, Stalin, C. S., Sagar, R., \& Wiita, P. J. 2003, ApJ, 586, L25
\bibitem[Gregg et al.(2000)]{gre00} Gregg, M. D., Becker, R. H., Brotherton, M. S., et al. 2000, ApJ, 544, 142 
\bibitem[Gregory et al.(1996)]{gre96} Gregory, P. C., Scott, W. K., Douglas, K., \& Condon, J. J. 1996, ApJS, 103, 427
\bibitem[Grandi \& Palumbo(2004)]{gra04} Grandi, P., \& Palumbo, G. 2004, Science, 306, 998 
\bibitem[Griffith \& Wright(1993)]{gri93} Griffith, M. R., \& Wright, A. E. 1993, AJ, 105, 1666
\bibitem[Griffith et al. (1995)]{gri95} Griffith, M. R., Wright, A. E., Burke, B. F., \& Ekers, R. D. 1995, ApJS, 97, 347
\bibitem[Gu et al. (2006)]{gu06} Gu, M. F., Lee, C.-U., Pak, S., Yim, H. S., \& Fletcher, A. B. 2006, A\&A, 450, 39
\bibitem[Gu \& Ai(2011)]{gu11} Gu, M. F., \& Ai, Y. L. 2011, A\&A, 528, A95
\bibitem[Gupta \& Joshi(2005)]{gup05} Gupta, A. C., \& Joshi, U. C. 2005, A\&A, 440, 855
\bibitem[Hu et al.(2006)]{hu06} Hu, S. M., Zhao, G., Guo, H. Y., Zhang, X., \& Zheng, Y. G. 2006, MNRAS, 371, 1243
\bibitem[Ivezi\'{c} et al.(2002)]{ive02} Ivezi\'{c}, \v{Z}., Menou, K., Knapp, G. R., et al. 2002, AJ, 124, 2364
\bibitem[Jiang \& Wang(2003)]{jia03} Jiang, D. R., \& Wang, T. G. 2003, A\&A, 397, L13 
\bibitem[Jorstad et al.(2007)]{jor07} Jorstad, S. G., Marscher, A. P., Stevens, J. A., et al. 2007, AJ, 134, 799
\bibitem[Joshi et al.(2011)]{jos11} Joshi, R., Chand, H., Gupta, A. C., \& Wiita, P. J. 2011, MNRAS, 412, 2717
\bibitem[Kimball \& Ivezi\'{c}(2008)]{kim08} Kimball, A.~E., \& Ivezi\'{c}, \v{Z}. 2008, AJ, 136, 684
\bibitem[Kong et al. (2006)]{kong06} Kong, M.~Z., Wu, X.~B., Wang, R., \& Han, J.~L. 2006, Chinese J. Astron.
Astrophys., 6,396
\bibitem[Li \& Cao(2008)]{li08} Li, S. L., \& Cao, X. W. 2008, MNRAS, 387, L41
\bibitem[Liu et al.(2006)]{liu06} Liu, Y., Jiang, D. R., \& Gu, M. F. 2006, ApJ, 637, 669
\bibitem[Liu et al. (2008)]{liu08} Liu, Y., Jiang, D. R., Wang, T. G., \& Xie, F. G. 2008, MNRAS, 391, 246
\bibitem[Lu et al.(2007)]{lu07} Lu, Y., Wang, T., Zhou, H., \& Wu, J. 2007, AJ, 133, 1615
\bibitem[Lundgren et al.(2007)]{lun07} Lundgren, B. F., Wilhite, B. C., Brunner, R. J., et al. 2007, ApJ, 656, 73
\bibitem[Lupton et al.(2002)]{lup02} Lupton, R. H., Ivezi\'{c}, \v{Z}., Gunn, J. E., et al. 2002, Proc. SPIE, 4836, 350
\bibitem[Massaro et al.(1998)]{mas98}Massaro, E., Nesci, R., Maesano, M., Montagni, F., \& D'Alessio, F. 1998, MNRAS, 299, 47
\bibitem[Montenegro-Montes et al.(2009)]{mon09} Montenegro-Montes, F. M., Mack, K.-H., Benn, C. R., et al.  2009, Astron. Nachr., 330, 157
\bibitem[Netzer(1990)]{Netzer90} Netzer, H. 1990 in Active Galactic Nuclei, ed. R. D. Blandford et al. (Berlin: Springer), 57
\bibitem[Peterson(1993)]{pet93} Peterson, B. M. 1993, PASP, 105, 247
\bibitem[Pian et al.(1999)]{pia99} Pian, E., Urry, C. M., Maraschi, L., et al. 1999, ApJ, 521, 112 
\bibitem[Poon et al.(2009)]{poo09} Poon, H., Fan, J. H., \& Fu, J. N. 2009, ApJS, 185, 511
\bibitem[Raiteri et al.(2001)]{rai01} Raiteri, C. M., Villata, M., Aller, H. D., et al. 2001, A\&A, 377, 396
\bibitem[Raiteri et al.(2007)]{rai07} Raiteri, C. M., Villata, M., Larionov, V. M., et al. 2007, A\&A, 473, 819
\bibitem[Rani et al.(2010)]{ran10} Rani, B., Gupta, A. C., Strigachev, A., et al. 2010, MNRAS, 404, 1992
\bibitem[Schneider et al. (2010)]{sch10} Schneider, D. P., Richards, G. T., Hall, P. B., et al. 2010, AJ, 139, 2360
\bibitem[Shang et al. (2005)]{sha05} Shang, Z. H., Brotherton, M. S., Green, R. F., et al. 2005, ApJ, 619, 41
\bibitem[Shen et al. (2011)]{she11} Shen, Y., Richards, G. T., Strauss, M. A., et al. 2011, ApJS, 194, 45 
\bibitem[Stalin et al.(2004)]{sta04} Stalin, C. S., Gopal-Krishna, Sagar, R., \& Wiita, P. J. 2004, MNRAS, 350, 175
\bibitem[Stalin et al.(2005)]{sta05} Stalin, C. S., Gupta, A. C., Gopal-Krishna, Wiita, P. J., \&  Sagar, R. 2005, MNRAS, 356, 607
\bibitem[Trump et al.(2006)]{tru06} Trump, J. R., Hall, P. B., Reichard, T. A., et al. 2006, ApJS, 165, 1
\bibitem[Turnshek(1988)]{tur88} Turnshek, D. A. 1988, in QSO absorption lines: probing the Universe, ed. J. C. Blades et al. (Cambridge), 17
\bibitem[Vagnetti et al.(2003)]{vag03} Vagnetti, F., Trevese, D., \& Nesci, R. 2003, ApJ, 590, 123
\bibitem[Vestergaard \& Peterson(2006)]{Vestergaard06} Vestergaard, M., \& Peterson, B.~M. 2006, ApJ, 641, 689
\bibitem[Villata et al.(2002)]{vil02}Villata, M., Raiteri, C. M., Kurtanidze, O. M., et al. 2002, A\&A, 390, 407
\bibitem[Wiita(1996)]{wii96} Wiita, P. J., 1996, in Miller H. R., Webb J. R., Noble J. C., eds, ASP Conf. Ser. Vol. 110, Blazar Continuum Variability. Astron. Soc. Pac., San Francisco, 42
\bibitem[Wu et al.(2005)]{wu05} Wu, J. H., Peng, B., Zhou, X., et al. 2005, AJ, 129, 1818
\bibitem[Wu et al.(2007)]{wu07} Wu, J. H., Zhou, X., Ma, J., et al. 2007, AJ, 133, 1599



\end{thebibliography}
\end{document}